\documentclass[preprint,proceedings]{rmaa}

\usepackage{epsf}
\usepackage{paralist}
\usepackage{psfrag,color}

\usepackage{lscape}
\usepackage{graphicx}

\newcounter{dddr}
\SetYear{2007}

\SetConfTitle{Proceedings of the Huatulco meeting on AGN}

\title{The Nature of the Far-UV Break\\
in the Energy Distribution of Quasars}

\author{Luc Binette,\altaffilmark{1}
Sinhu\'e Haro-Corzo,\altaffilmark{2}
Yair Krongold,\altaffilmark{1}
and Anja C. Andersen\altaffilmark{3} }

\altaffiltext{1}{Instituto de Astronom\'\i{}a, UNAM, M\'exico D.F.,
M\'exico}

\altaffiltext{2}{Instituto de Ciencias Nucleares, UNAM, M\'exico
D.F.}

\altaffiltext{3}{Dark Cosmology Center, Copenhagen, Denmark}

\shortauthor{Binette et\,al.}

\shorttitle{The far-UV Break of Quasars}

\fulladdresses{
\item  Luc Binette, Sinhu\'e Haro-Corzo, Yair Krongold: Instituto de
Astronom\'\i{}a, Universidad Nacional Aut\'onoma de M\'exico, Ciudad
Universitaria, Apartado Postal 70--264, CP 04510, M\'exico D.F.,
M\'exico.
\item Anja C. Andersen: Dark Cosmology Center, Juliane Maries Vej 30,
DK-2100 Copenhagen, Denmark. }

\listofauthors{L. Binette,  S. Haro-Corzo, Y. Krongold, A. C.
Andersen}

\indexauthor{Binette, L.}

\indexauthor{Haro-Corzo, S.}

\indexauthor{Krongold, Y.}

\indexauthor{Andersen, A. C.}

\abstract{A prominent continuum steepening is observed in quasar
energy distributions near 1100\AA. We review possible
interpretations for the physical origin of this so-called far-UV
break.}

\resumen{Se observa un fuerte empinamiento  en la distribuci\'on de
energ\'\i a espectral de los cuas\'ares alrededor de 1100\AA.
Revisamos posibles interpretaciones  para el orig\'en f\'{\i}sico de
este quiebre en el ultravioleta lejano. }

\addkeyword{ISM: dust }
\addkeyword{galaxies: active}
\addkeyword{radiative transfer}
\addkeyword{ultraviolet: general}

\begin{document}

\maketitle


\section{Introduction}

\begin{figure}[!ht]
\begin{center}
\includegraphics[width=8.0cm,keepaspectratio=true]{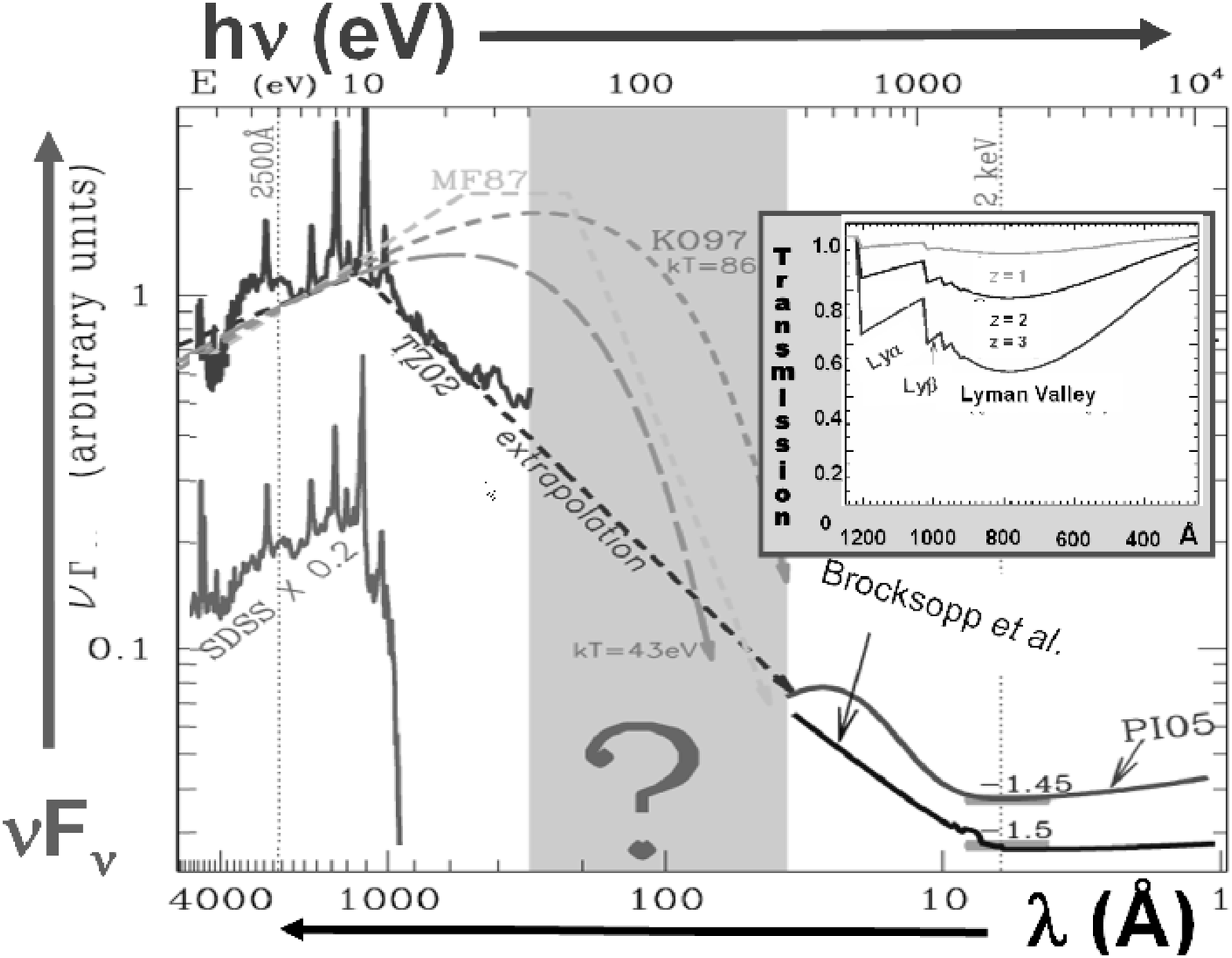}
\caption{Figure showing the UV composite spectra for quasars derived
by TZ02 and SDSS (Vanden Berk et\,al. 2001) (the SDSS curve is moved
down by 0.7\,dex to avoid cluttering). The individual quasar spectra
used in the TZ02 composite were divided by the transmission function
(see inset) appropriate for their redshift before being averaged.
The SDSS composite did not consider such a correction. Also shown is
are the averages of two series of X-ray models that fit individual
XMM-Newton spectra. They correspond to the powerlaw+blackbody fits
of Piconcelli et\,al. (2005: PI05) and to the broken power-law fits
of Brocksopp et\,al. (2006), respectively. The vertical positions of
these X-ray composites were arbitrarily set to reproduce an
$\alpha_{OX}$ of $-1.45$ and $-1.50$, respectively. If, instead of
composites, we considered SEDs from \emph{individual} quasars, it
could be shown that the extrapolation of the far-UV does \emph{not}
connect smoothly with the soft-X-ray excess component (see H07 for
further details). \label{fig:sed}}
\end{center}
\end{figure}

The spectra of quasars and Seyfert galaxies show strong emission
lines superimposed onto a bright continuum.  The continuum contains
a significant feature in the optical-ultraviolet region, known as
``the Big Blue Bump'' (BBB).  Full coverage of the BBB region can
only be gathered from significantly redshifted objects, since the
Galaxy is opaque to the Lyman continuum.  Although it has been
possible to infer the UV Spectral Energy Distribution (SED) of
quasars down to $\sim$350\AA, it remains poorly known between the
extreme UV and  the X-rays. In the X-rays, a soft excess component
is reported (Fig.\,\ref{fig:sed}), which cannot be reduced to a
simple extrapolation of the far-UV powerlaw [Haro-Corzo (this
meeting) and Haro-Corzo et\,al. (2007: H07)].

Composite energy distributions of the BBB were derived by Zheng
et\,al. (1997) and later by Telfer et\,al. (2002: TZ02) (reproduced
in Fig.\,\ref{fig:sed}) using archived HST-FOS spectra. Before
averaging, each spectrum was dereddened for Galactic absorption as
well as statistically corrected for the absorption due to
intergalactic Ly$\alpha$ absorbers and Lyman limit systems. The
transmission function for intergalactic absorption is illustrated in
the inset of Fig.\,\ref{fig:sed} for redshifts 1, 2 and 3. Without
such a correction, the far-UV continuum appears suppressed, as shown
by the SDSS composite.

A striking feature of the composite quasar SED of TZ02 (see
Fig.\,\ref{fig:sed}) is that a significant steepening occurs around
1100\AA, leading to a far-UV powerlaw of index $\nu^{-1.7}$
($F_{\nu}\propto \nu^{+\alpha}$). Such a steepening of the continuum
is present not only in composite SEDs, but is also evident  in
individual quasar spectra such as in PG\,1148+549
(Fig.\,\ref{fig:disk}) (see other examples in Binette et\,al.2005;
hereafter B05). The degree of steepening also varies significantly
from object to object\footnote{Scott et\,al. (2004) reported a lack
of evidence of a far-UV break in the case of nearby, less luminous
AGN.}. We will refer to this continuum steepening as the far-UV
break. Korista, Ferland \& Baldwin (1997a; KO97) pointed out the
difficulties of reproducing the equivalent widths of the high
ionization lines of He{\sc ii} $\lambda$1640\AA, C{\sc iv}
$\lambda$1549\AA\ and O{\sc vi} $\lambda$1035\AA, assuming a
powerlaw as soft as $\nu^{-2}$. State of the art photoionization
models favor a much harder SED, one that peaks in the extreme-UV
beyond 20\,eV. In Fig.\,\ref{fig:sed}, we show two theoretical SEDs
characterized by an exponential cut-off of 43 and 86\,eV. These were
used in the BELR models of Korista et\,al. 1997b (see also Baldwin
et\,al. 1995; Casebeer, Leighly \& Baron 2006). We will refer to
this incongruity between the theoretical SEDs favored by
photoionization and the steep far-UV SEDs actually observed as the
`softness problem'.

\begin{figure}[!ht]
\includegraphics[width=7.0cm,keepaspectratio=true]{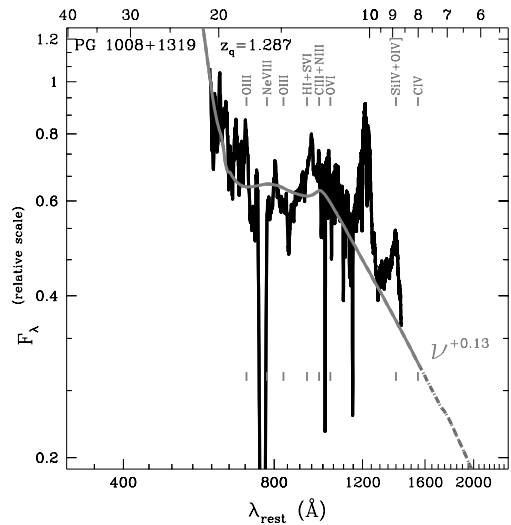}
\caption{Rest-frame spectrum of PG\,1008+1319 (thin black line). The
thick grey line represents an absorption model of a powerlaw SED,
assuming extinction by cubic diamonds (using extinction curve D3 in
H07). \label{fig:pgten}}
\end{figure}

\begin{figure}[!ht]
\includegraphics[width=7.4cm,keepaspectratio=true]{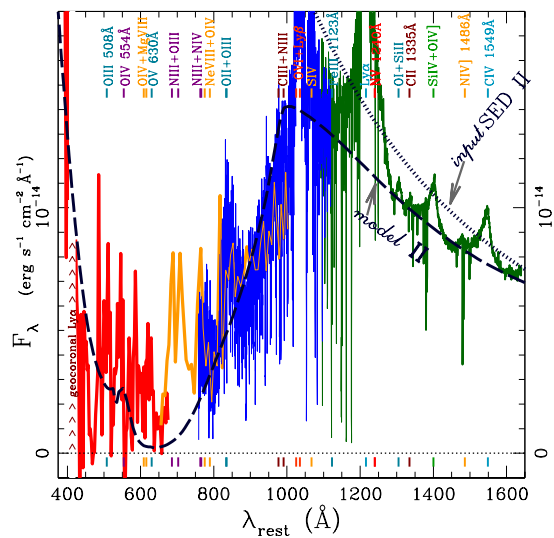}
\caption{Rest-frame spectrum of Ton\,34 as derived from archive
spectra from IUE, HST and Palomar (Sargent et\,al. 1988). The dash
line represents an absorption model of the SED represented by the
dotted line, assuming extinction by cubic nanodiamonds.
\label{fig:ton}}
\end{figure}

In our study of this problem, we have explored the shape of the
far-UV break and whether it is followed by an upturn in the extreme
UV, allowing the intrinsic SED to be much harder beyond 30\,eV than
suggested by an extrapolation of the far-UV continuum. If such an
upturn existed, it would likely solve the softness problem
identified above. In order to be able to evaluate the breath of the
continuum break, that is, to identify a possible upturn in the
extreme UV, we found that this was feasible using HST-FOS spectra,
provided the quasars were of redshift $\simeq 1$ and  had been
observed with more than one grating. These conditions were met only
for a few objects within the TZ02 sample that B05 analyzed. They
reported an upturn in four quasars. One example, is PG\,1008+1319 at
$z=1.287$ (Fig.\,\ref{fig:pgten}). After combining various archival
spectra, Binette \& Krongold (in preparation) reported a hint of a
possible far-UV rise in a fifth object, Ton\,34 ($z=1.928$) (see
Fig.\,\ref{fig:ton}). This object is characterized by the largest
steepening known ($F_{\nu} \propto \nu^{-5.3}$ in the far-UV). A
sixth example is provided by the well studied quasar HE\,2347$-$4342
($z=2.885$), where an upturn is visible shortward of 700\AA\ as
shown in Fig.\,\ref{fig:hei}.

There is no generally accepted interpretation of the nature of the
far-UV break. Below, in \S\,\ref{sec:cau}, we review possible
absorption mechanisms that would give rise to the break and
furthermore allow the emergence of an upturn in the extreme-UV, in
order to solve the `softness problem' defined above.


\section{Possible causes for the UV-break} \label{sec:cau}

\begin{figure}[!ht]
\begin{center}
\includegraphics[width=8.0cm,keepaspectratio=true]{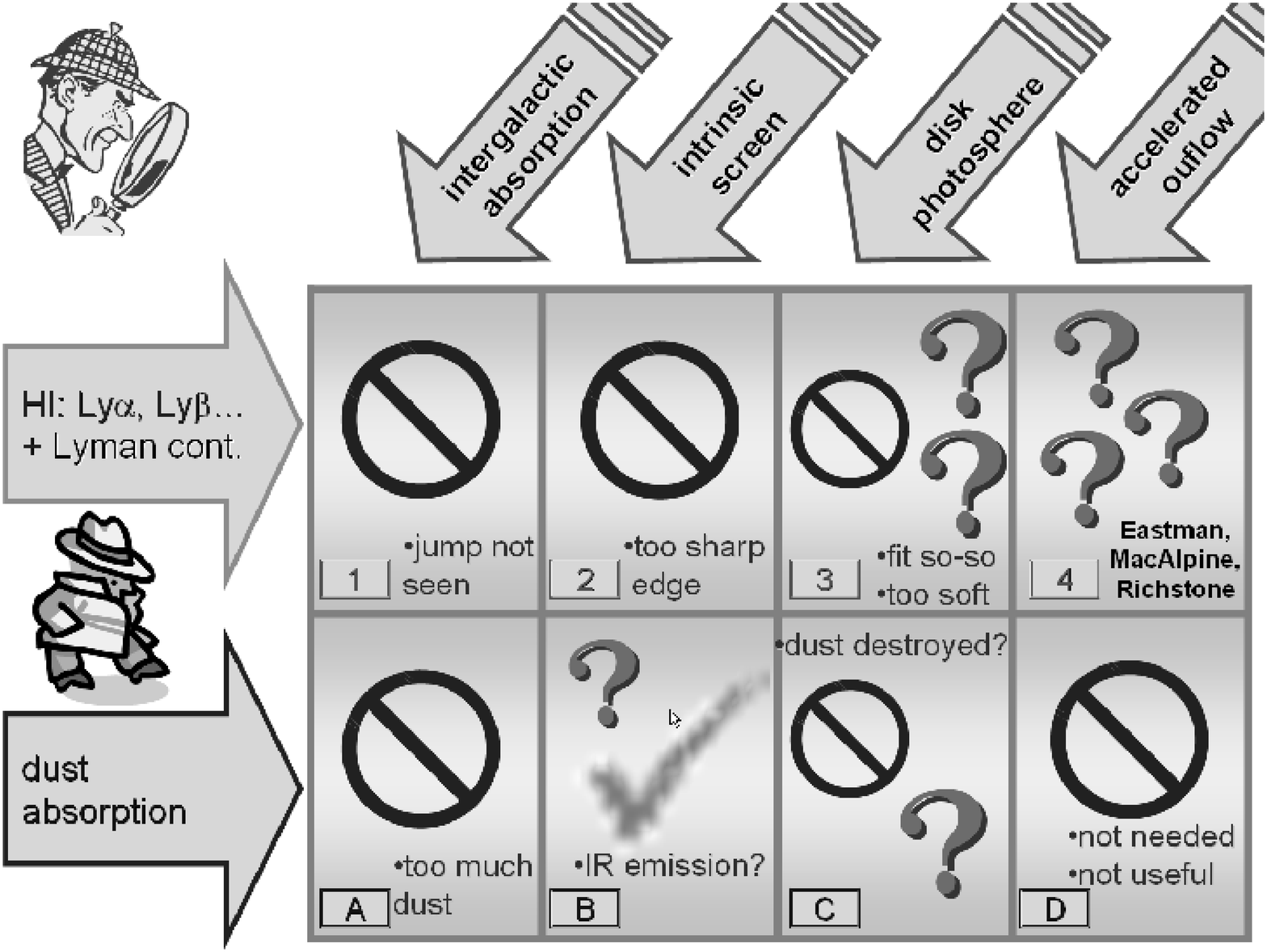}
\caption{Diagram  illustrating the 8 possible \emph{cases} arising
from either H{\sc i} or dust absorption. Barred circles denote
rejected or unpromising cases, while question marks suggest a
certain level of doubt or a conclusion that is possibly premature.
\label{fig:all}}
\end{center}
\end{figure}

We hereafter assume that the far-UV break results from absorption
and  will consider two possibilities: [I]-- H{\sc i} (Ly$\alpha$,
$\beta$, $\gamma$ ... + bound-free) and [II]-- dust. We will
consider four locations for the absorbing medium: ($i$)
intergalactic, ($ii$) local to the quasar ISM, ($iii$) accretion
disk photosphere and ($iv$) accelerated outflow. The resulting eight
cases are illustrated in Fig.\,\ref{fig:all} with labels 1--4 for
H{\sc i} and A--D for the dust. Among the eight cases reviewed
below, many impact positively the softness problem, either because
the local BELR sees a different SED than the observer (e.g.
intergalactic absorption), or because the absorption in the UV is
followed by a flux upturn at higher energies.

\begin{figure}[!ht]
\includegraphics[width=7.8cm,keepaspectratio=true]{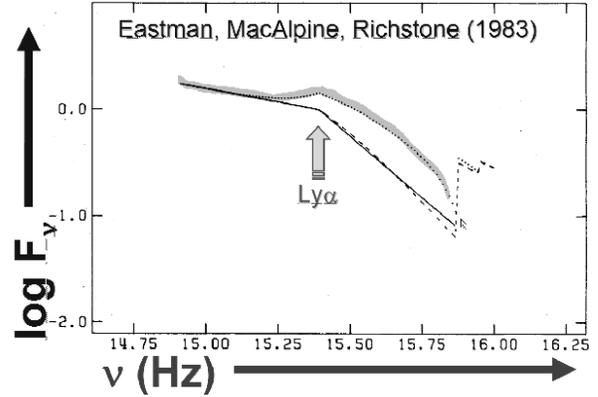}
\caption{The accelerated cloud absorption model (top grayed line)
proposed by Eastman, MacAlpine \& Richstone (1983).
\label{fig:east}}
\end{figure}

\begin{list}{Case~\arabic{dddr} --}{\usecounter{dddr}\setcounter{dddr}{0}}
\item Binette et\,al. (2003) studied the possibility of intergalactic H{\sc i} absorption. They assumed
a behavior of the H{\sc i} density as a function of $z$ proportional
to the gas density expected from the warm-hot intergalactic medium.
Although they could reproduce the steepening observed in the TZ02
composite, they rejected this possibility, since it implied
 a continuum flux  discontinuity  at 1216\AA\
(observer-frame) that is not observed.

\item An intrinsic absorber at a redshift close to the quasar results in a sharp
absorption edge near 912\AA\ as well as in a saturated  Ly$\alpha$
absorption line. A sharp edge at 912\AA\ has been reported in only
$\sim$10\% of AGN (see KO97) and is unrelated to the far-UV break
discussed in this paper, which consists in a change in spectral
index rather than a discontinuous edge. The situation described by
case\,2 is therefore not relevant to our study of the UV break.

\begin{figure}[!ht]
\includegraphics[width=7.5cm,keepaspectratio=true]{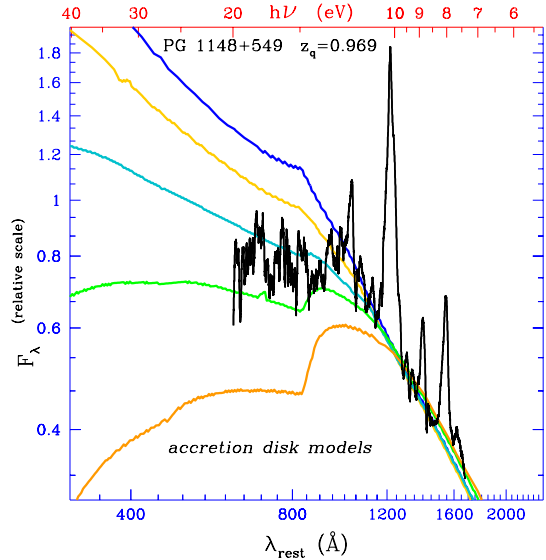}
\caption{The thin black line represents the far-UV spectrum of
PG\,1148+549. Five accretion disk models from Hubeny et\,al. (2000)
are shown for comparison assuming a (face-on) stationary disk
structure and a maximum rotating Kerr blackhole of mass
$10^{9}\,$M$_{\sun}$, with accretion rates of 2, 1, 1/2, 1/4 and
1/8\,M$_{\sun}$\,yr$^{-1}$, respectively. \label{fig:disk}}
\end{figure}

\item State of the art models of `bare' accretion disks predict
a steepening (i.e. a Lyman edge) near Ly$\alpha$ for a certain range
of accretion rates and black hole masses. This is illustrated in
Fig.\,\ref{fig:disk}, where we show  different SED models from
Hubeny et\,al. (2000) that include a detailed non-LTE treatment of
abundant elements as well as continuum opacities due to bound-free
and free-free transitions. We find the fit to the far-UV break in
PG\,1148+549 to be unsatisfactory. A more important aspect to
consider is that disk models that produce a break near the Lyman
edge are not followed by a flux upturn at higher energies. Hence the
softness problem remains whole with current disk models that intend
to reproduce the break. A similar shortage of hard photons also
characterizes the comptonized accretion disk model proposed by Zheng
et\,al. (1997), even though in that case the UV break is much better
reproduced.  We may conjecture that current stationary disks fail,
because they do not include thermal inhomogeneities, Parker
instabilities or other energy transport mechanisms (Blaes 2007).
Therefore, we cannot rule out that the next generation of disk
models will succeed to simultaneously be very efficient in the
far-UV and yet produce a trough near the Lyman limit. Both of these
features are required to fit the SED of HE\,2347$-$4342
(Fig.\,\ref{fig:hei}). Disks  are also expected to generate a wind
(see Proga 2007a, and references therein). Because a reduction of
the surface temperature might be expected at the point where the
wind is launched as a result of reduced viscous heating, we may
speculate that a narrow disk wind (Proga 2007b), if launched from
the position where the bulk of the 1000\AA\ flux originates, might
cause an absorption trough where the break is observed.

\begin{figure}[!ht]
\includegraphics[width=7.5cm,keepaspectratio=true]{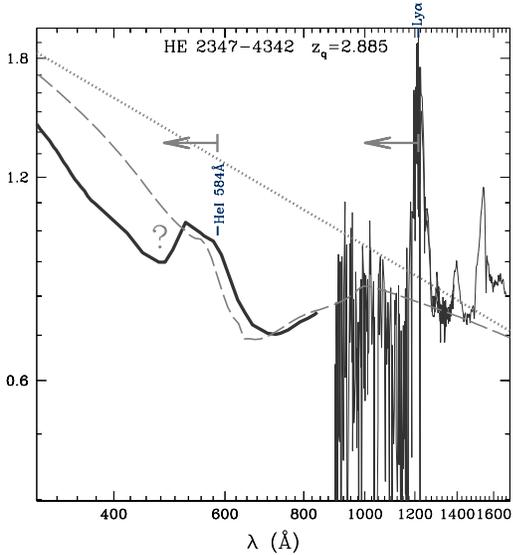}
\caption{Spectrum of HE\,2347$-$4342 (thick and thin black lines)
extracted from Reimers et\,al. (1998). The dashed line represents an
absorption model of the SED represented by the dotted line, assuming
extinction by cubic nanodiamonds. \label{fig:hei}}
\end{figure}

\item Eastman, MacAlpine \& Richstone (1983) produced a steepening of the
continuum at 1200\AA\ by integrating the absorption from clouds
progressively accelerated up to 0.8$c$. Such a model, after
inclusion of  cloud emission, is represented by the thick grayed
line in Fig.\,\ref{fig:east}. A more thorough exploration of this
type of model would be needed to determine whether it is  a
promising avenue. Would a wind structure rather than discrete clouds
not be preferable? A different behavior of the H{\sc i} opacity with
velocity might be able to shift the break position to the observed
value near 1100\AA\ (instead of $\simeq 1216$\AA). The upturn in
flux beyond 20\,eV also remains to be calculated. If He{\sc i}
opacity were included, such models might succeed in explaining the
odd dip observed at 500\AA\ in HE\,2347$-$4342, one of the most
studied quasar (see Fig.\,\ref{fig:hei}). Interestingly, both breaks
at 1100\AA\ and 500\AA\  are  blueshifted  with respect to
rest-frame Ly$\alpha$ from He{\sc i} and H{\sc i}, at 584\AA\ and
1216\AA, respectively.

\end{list}

\begin{list}{Case~\Alph{dddr} --}{\usecounter{dddr}\setcounter{dddr}{0}}

\item  B05 studied the possibility of absorption by intergalactic
dust consisting of nanodiamond grains (other dust compositions were
unsuccessful). They could reproduce the wide dip centered on
$z\simeq 1$ displayed by the far-UV indices $\alpha_{FUV}$ when
plotted with redshift (Fig.\,12 in B05). This possibility was
rejected, since it required too much intergalactic dust ($\sim$17\%
of all cosmic carbon!).

\item Shang et\,al. (2005) previously explored the
possibility of ISM and SMC-like extinction to explain the break.
They rejected this hypothesis, because the absorption resulted in a
broad curvature for the SED, which is not observed. Assuming a dust
component composed of crystallite carbon grains, B05 successfully
reproduced the far-UV break observed in individual quasars of the
TZ02 sample. The case of PG\,1008+1319 is illustrated in
Fig.\,\ref{fig:pgten}. The softness problem disappears, since the
dust becomes transparent at higher energies and a flux upturn is
therefore predicted below 650\AA. This allows the intrinsic SED to
be much harder in the extreme UV than the flux measured next to the
break suggests. Two other examples of an upturn in the extreme UV
are presented here: Ton\,34 (Fig.\,\ref{fig:ton}) and
HE\,2347$-$4342 (Fig.\,\ref{fig:hei}).

\item B05 found that the column of dust required to fit the far-UV break
was about the same for the majority of their quasars. This is an
improbable situation for the above case\,B, where the dust screen is
presumed independent of the continuum source. Such a narrow range in
dust opacities would make more sense if the dust screen was somehow
connected to the continuum source. What about the possibility of the
dust being part of the plasma where the UV is produced? Even though
nanodiamond formation is favored by the presence of UV
radiation\footnote{It is interesting to note that via the mid-IR
emission bands observed at 3.43 and 3.53$\mu$m (van Kerckhoven
et\,al. 2002), nanodiamonds have so far only been identified around
3 stellar disks that turn out to be heated by the UV radiation from
the central star.} (Duley \& Grishko 2001; Kouchi et\,al. 2005), the
high temperatures of the inner disk coupled with its high gas
density would certainly destroy the dust before it reaches the
radius where the UV is radiated. We can therefore rule out this
possibility. An interesting variant of case\,C that cannot so easily
be ruled out is provided by having the dust manufactured within the
much cooler outer regions of the disk and then launched as part of a
funneled disk wind. We may conjecture that this dusty wind would
later intercept our line-of-sight to the the inner UV-emitting
regions of the disk.

\item We consider the hypothesis of acceleration of
dusty material to \emph{high} velocities superfluous, for it is
neither needed nor useful for the purpose of reproducing the far-UV
break.

\end{list}

\section{Discussion}

Crystallite carbon dust has not been yet observed in emission in the
mid-IR. De\,Diego et\,al. (2007) actually ruled out meteoritic
nanodiamond emission in 3C298 using \emph{Spitzer} data. H07 has
modified the original model of B05 by discarding meteoritic
nanodiamonds and making use of cubic nanodiamonds only (i.e. gains
without surface adsorbates). Attempts will be made to detect
emission due to bulk impurities from this type of grains. A possible
problem of the dust screen model is the absence of absorption lines
associated with the dust screen (see Binette \& Krongold 2007).

To summarize, out of the 8 possibilities of Fig.\,\ref{fig:all}, we
favor case\,B as our overall favored model. This is possibly more a
reflection of this case having been a simpler paradigm to develop
and later closely compared with the spectra. We consider the
remaining three alternative cases 3, 4 and C as certainly worth
being explored further.

\acknowledgements This work was funded by the CONACyT grants J-49594
and J-50296, and the UNAM PAPIIT grant IN118905. The Dark Cosmology
Centre is funded by the Danish National Research Foundation.


\onecolumn


\begin{thebibliography}{}

\bibitem[]{} Baldwin, J., Ferland,
G., Korista, K., \& Verner, D.\ 1995, \apjl, 455, L119


\bibitem[Binette et al.(2003)]{bi03} Binette, L.,
Rodr\'{\i}guez-Mart\'{\i}nez, M., Haro-Corzo, S., \& Ballinas, I.\
2003, \apj, 590, 58

\bibitem[Binette et al.(2005)]{bm05} Binette, L.,  Magris C., G.,
Krongold, Y., Morisset, C., Haro-Corzo, S., de Diego, J. A.,
Mutschke, H., \& Andersen, A.\ 2005, \apj, 631, 661 (B05)

\bibitem[Binette \& Krongold(2007)]{bm07} Binette, L. \&
Krongold, Y.\ 2007, \apj, submitted





\bibitem[Blaes(2007)]{2007astro.ph..3589B} Blaes, O.\ 2007, in ASP Conf. Ser., ed. L.C. Ho and J.-M. Wang,
The central engine of active galactic nuclei, in press,
astro-ph/0703589


\bibitem[Brocksopp et al.(2006)]{2006MNRAS.366..953B} Brocksopp, C.,
et\,al. 2006, \mnras, 366, 953

\bibitem[Casebeer et al.(2006)]{2006ApJ...637..157C} Casebeer, D.~A.,
Leighly, K.~M., \& Baron, E.\ 2006, \apj, 637, 157

\bibitem[de\,Diego et al.(2007)]{diego07} de\,Diego, J.~A.,
Binette, L., Ogle, P., Andersen, A.~C., Haro Corzo, S., \& Wold, M.\
2007, \aap, {467}, L7

\bibitem[Duley \& Grishko(2001)]{2001ApJ...554L.209D} Duley, W.~W., \&
Grishko, V.~I.\ 2001, \apjl, 554, L209

\bibitem[Eastman et al.(1983)]{1983ApJ...275...53E} Eastman, R.~G.,
MacAlpine, G.~M., \& Richstone, D.~O.\ 1983, \apj, 275, 53


\bibitem[Haro-Corzo et al.(2007)]{haro07} Haro-Corzo, S., Binette, L.,
Krongold, Y., Benitez, E., Humphrey, A., Nicastro, F., \&
Rodriguez-Martinez, M., 2007, \apj, 662, 145  (H07)


\bibitem[Hubeny et al.(2000)]{2000ApJ...533..710H} Hubeny, I., Agol, E.,
Blaes, O., \& Krolik, J.~H.\ 2000, \apj, 533, 710


\bibitem[Korista et al.(1997a)]{1997ApJ...487..555K} Korista, K., Ferland,
G., \& Baldwin, J.\ 1997a, \apj, 487, 555 (KO97)

\bibitem[Korista et al.(1997b)]{1997ApJS..108..401K} Korista, K., Baldwin,
J., Ferland, G., \& Verner, D.\ 1997b, \apjs, 108, 401

\bibitem[Kouchi et al.(2005)]{2005ApJ...626L.129K} Kouchi, A.,
Nakano, H., Kimura, Y., \& Kaito, C.\ 2005, \apjl, 626, L129






\bibitem[Piconcelli et al.(2005)]{2005A&A...432...15P} Piconcelli, E.,
et\,al. 2005, \aap, 432, 15 (PI05)

\bibitem[Proga(2007a)]{2007astro.ph..1100P} Proga, D.\ 2007a,in ASP Conf. Ser., ed. L.C. Ho and J.-M. Wang,
The central engine of active galactic nuclei,  in press,
astro-ph/0701100

\bibitem[Proga(2007b)]{2007astro.ph..2582P} Proga, D.\ 2007b, \apj, in
press, astro-ph/0702582



\bibitem[Sargent et al.(1988)]{sargent88} Sargent, W.~L.~W.,
Boksenberg, A., \& Steidel, C.~C.\ 1988, \apjs, 68, 539

\bibitem[Shang et al.(2005)]{2005ApJ...619...41S} Shang, Z., et al.\ 2005,
\apj, 619, 41

\bibitem[Reimers et al.(1998)]{1998uabi.conf..579R} Reimers, D.,
K\"{o}hler, S., Hagen, H.-J., \& Wisotzki, L.\ 1998, ESA SP-413:
Ultraviolet Astrophysics Beyond the IUE Final Archive, 579


\bibitem[Scott et al.(2004)]{2004ApJ...615..135S} Scott, J.~E., Kriss,
G.~A., Brotherton, M., Green, R.~F., Hutchings, J., Shull, J.~M., \&
Zheng, W.\ 2004, \apj, 615, 135

\bibitem[Telfer et al.(2002)]{telfer} Telfer, R. C., Zheng, W., Kriss, G. A.,
Davidsen, A. F.\ 2002, \apj, 565, 773 (TZ02)

\bibitem[Vanden Berk et al.(2001)]{vandenberk01} Vanden Berk, D.~E.,
et al.\ 2001, \aj, 122, 549

\bibitem[van Kerckhoven et al.(2002)]{kerckhoven} van Kerckhoven, C.,
Tielens, A. G. G. M., Waelkens, C.\ 2002, \aap, 384, 568


\bibitem[Zheng et al.(1997)]{1997ApJ...475..469Z} Zheng, W., Kriss, G.~A.,
Telfer, R.~C., Grimes, J.~P., \& Davidsen, A.~F.\ 1997, \apj, 475, 469


\end{thebibliography}
\end{document}